%% file: root.tex
\documentclass{ifacconf}

\usepackage{graphicx, subfigure}
\usepackage{natbib}        
\usepackage{units}
\usepackage{verbatim}
\usepackage[normalem]{ulem}
\usepackage{textcomp}
\usepackage{gensymb}
\usepackage{verbatim}
\usepackage{url}
\makeatletter
\let\algorithm\@undefined
\let\endalgorithm\@undefined
\makeatother
\usepackage[ruled,linesnumbered]{algorithm2e}	
\usepackage{pgfplots}
\usepackage{tikz}
\makeatletter
\let\theoremstyle\@undefined
\let\case\@undefined
\let\c@case\@undefined
\makeatother
\usepackage{amsmath,amssymb,amsfonts}

\usepackage[acronym,nomain]{glossaries}
\definecolor{darkgreen}{rgb}{0.0, 0.5, 0.0}
\usetikzlibrary{fit, shapes}

\newtheorem{lemma}{Lemma}
\newtheorem{case}{Case}

\begin{document}
\begin{frontmatter}

\title{Vision-Based Real-Time Indoor Positioning System for Multiple Vehicles \thanksref{footnoteinfo}} 

\thanks[footnoteinfo]{This research is supported by the Deutsche Forschungsgemeinschaft (DFG, German Research Foundation) within the Priority Program
SPP 1835 “Cooperative Interacting Automobiles” and the Post Graduate Program GRK 1856 “Integrated Energy Supply Modules for Roadbound E-Mobility”.}

\author[First]{Maximilian Kloock} 
\author[First]{Patrick Scheffe} 
\author[First]{Isabelle T\"ulleners}
\author[First]{Janis Maczijewski}
\author[First]{Stefan Kowalewski}
\author[First]{Bassam Alrifaee}

\address[First]{Chair for Embedded Software, RWTH Aachen University, \\52074 Aachen, Germany (corresponding e-mail: kloock@embedded.rwth-aachen.de).}

\newacronym{RFID}{RFID}{Radio Frequent Identification}
\newacronym{GPS}{GPS}{Global Positioning System}
\newacronym{GNSS}{GNSS}{Global Navigation Satellite System}
\newacronym{IPS}{IPS}{Indoor Positioning System}
\newacronym{RF}{RF}{Radio Frequency}
\newacronym{pnp}{PnP}{Perspective from $n$ Points}
\newacronym{IMU} {IMU} {Inertial Measurement Unit}
\newacronym{DDS} {DDS}{Data Distribution Service}

\newacronym{SLAM} {SLAM}{Simultaneous Localization and Mapping}
\newacronym{SfM} {SfM}{Structure from Motion}
\newacronym{VLC} {VLC}{Visible Light Communication}

\newacronym{pose} {pose}{position and orientation}
\newacronym{CPM Lab}{CPM Lab}{Cyber-Physical Mobility Lab}

\begin{abstract}                
We propose a novel external indoor positioning system that computes the position and orientation of multiple model-scale vehicles. For this purpose, we use a camera mounted at a height of \unit[3.3]{m} and LEDs attached to each vehicle. We reach an accuracy of about \unit[1.1]{cm} for the position and around \unit[0.6]{\degree} for the orientation in the mean. Our system is real-time capable with a soft deadline of \unit[20]{ms}. Moreover, it is robust against changing lighting conditions and reflections.
\end{abstract}


\end{frontmatter}
\section*{Supplementary material}
A demonstration video of this work is available at \\ \url{https://youtu.be/k6aD5G9DW4o}. 

More information about the CPM Lab is provided at \\ \url{https://cpm.embedded.rwth-aachen.de} 

\section{Introduction}
Applications for indoor positioning are, e.g., humanoid robots or (model-scale) autonomous vehicles. We are building the \gls{CPM Lab} with 20 model-scale vehicles ($\mu$Cars) to develop and to evaluate algorithms for networked and autonomous vehicles. For this purpose, we developed an \gls{IPS}, since the knowledge of the vehicle positions is significant as computations of trajectories and the interaction between vehicles highly depends on their current positions. Additionally, trajectory control depends on the accuracy of positioning. The precision of the \gls{IPS} is improved by sensor fusion with dead reckoning data. Nevertheless, the \gls{IPS} is the only absolute reference system for the positioning. Therefore, to test functionalities in the field of networked and autonomous vehicles using model-scale vehicles, an accurate \gls{IPS} is required.\\
There are already many systems providing the position of indoor robots. Some approaches are wave-based using WLAN or \gls{RFID}, e.g., \citep{wireless3, RFID2}, or ultrasonic and \gls{RF}, e.g., \citep{DissDiab,wireless}. Overall, those approaches have the deficiency of an accuracy in the decimeter to meter range. More precise approaches are vision-based.\\
There are many different attempts for vision-based indoor positioning, e.g.,~\citep{optical}. Some use feature detection, e.g., \citep{Feature1, InfraredAreas}, some use \gls{VLC}, e.g., \citep{VLC1, VLC2, VLC3} and others use \gls{SLAM} methods, e.g., \citep{SLAM2, SLAM4}. However in those approaches the target object is equipped with a camera and positions itself depending on its view. This requires additional computation power on the robots. In order to keep the computation requirements on the vehicles low, we develop a system which externally determines the position and the orientation (pose) of the target object. Our approach is inline with the attempt of \citep{InfraredLeds}. However, \citep{InfraredLeds} deals only with a single vehicle for positioning. Our \gls{IPS} is able to determine the pose of multiple vehicles. \\ 
There are also approaches to detect tags, e.g. \citep{tags, tags3}. However, due to changing lighting conditions, the tags have to be larger than the model-scale vehicles. Therefore, such approaches are not suitable in our case. \\
We propose a new vision-based \gls{IPS} that externally computes the poses of multiple model-scale vehicles. For this, we use LEDs attached to the autonomous vehicles and a camera mounted on the ceiling. The LEDs can be detected robustly even with changing lighting conditions using short exposure times and bright LEDs. To distinguish the vehicles, one of the LEDs flashes in a vehicle specific frequency. Furthermore, our system can be used in a real-time environment with a soft deadline of \unit[20]{ms}.

The rest of the paper is structured as follows. Section~\ref{sec:3} gives an overview of the infrastructure and the \gls{IPS} algorithm. The correctness of this algorithm is shown in Section~\ref{sec:4}. Section~\ref{sec:evaluation} evaluates our \gls{IPS} and Section~\ref{sec:conclusion} concludes this paper.

\section{INDOOR POSITIONING SYSTEM}
\label{sec:3}

\subsection{System Overview}

\begin{figure}
\centering
\includegraphics[width=0.8\columnwidth]{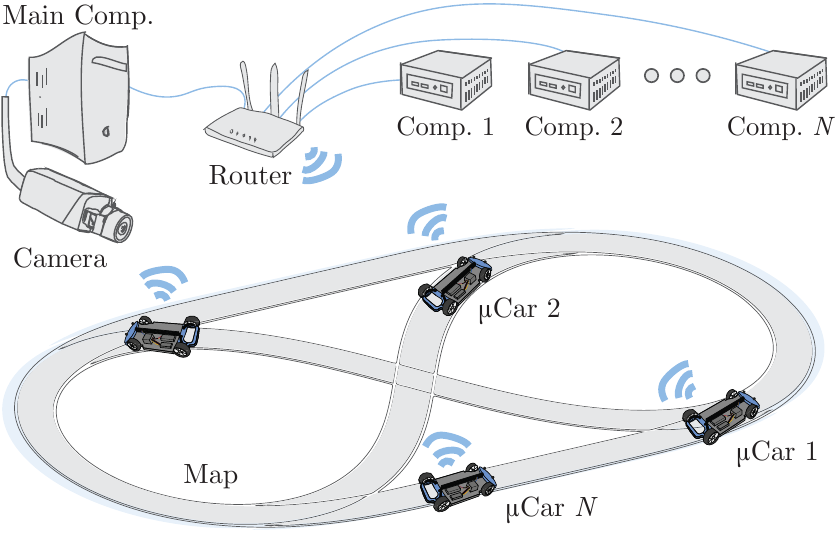}
\caption{System overview. $\mu$Cars drive on a printed map and control trajectories planned by external computation units in a networked manner. The camera is used for positioning of the vehicles.}
\label{Fig:overview}
\end{figure}

Fig.~\ref{Fig:overview} sketches our system overview. Up to 20 vehicles drive on a map of 4mx4.5m size. One external computation device is provided for each vehicle. A camera records the whole map and a main computer performs the image processing. A router allows for wireless communications between the external computation devices, the main computer, and the vehicles. For more details of the \gls{CPM Lab}, see \citep{kloock2020cyber}. 

\begin{figure}
\centering
\includegraphics[width=0.75\columnwidth]{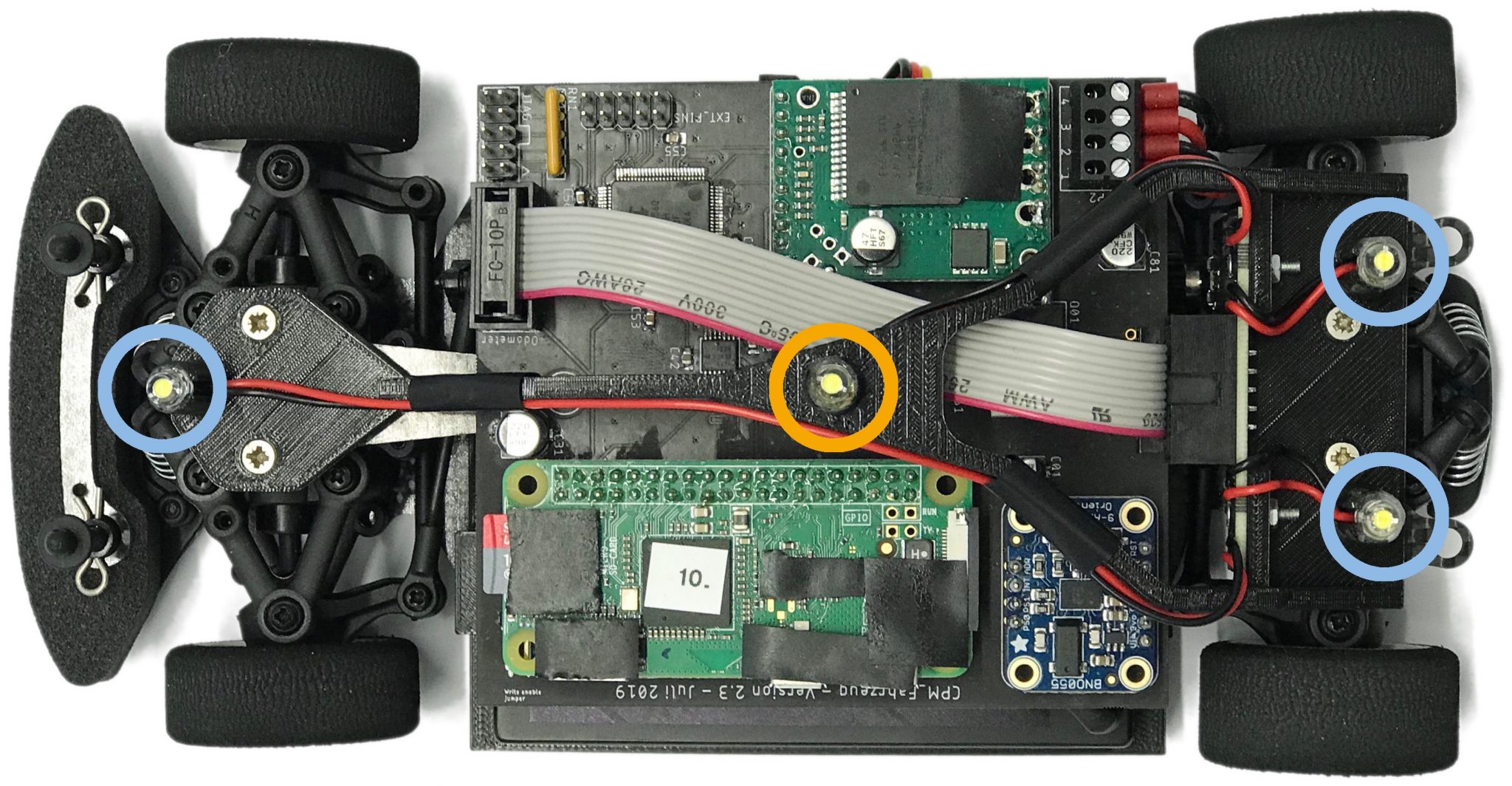}
\caption{$\mu$Car with four mounted LEDs. The blue-marked LEDs (outer ones) are the positioning LEDs where the yellow-marked LED (inner one) describes the identification LED.}
\label{Fig:Vehicle}
\end{figure}

We mount the camera at a height of 3.3 meters straight downwards. Its field of view covers the whole map. Furthermore, we attach four LEDs to each vehicle as shown in Fig.~\ref{Fig:Vehicle}. The three outer LEDs marked in blue are arranged in a non-equilateral triangle and used to determine the pose of the vehicle. Since the triangle is not equilateral, the direction of the vehicle can be depicted. We use the fourth LED marked in yellow to distinguish the vehicles. For this purpose, it flashes in a vehicle specific frequency. This LED lays in the borders of the triangle of the outer LEDs to simplify the clustering of the points to a vehicle. Since a frequency cannot be determined in a single image, we consider a sequence of images. However for pose determination, we only use the latest one. For more details of the vehicle construction, see~\citep{scheffe2020}. 

\subsection{Algorithm Overview}

Our algorithm consists of five steps. In a first one, a camera stream is received. In the second step, we extract the LED points.  For each image, we cluster the detected points belonging to a single vehicle afterwards. In a fourth step, the detected vehicles in the current image are mapped to the vehicles from the previous one. With this, we gain the sequence of points belonging to each vehicle in different images. Having this sequence, we compute the pose and the ID of each vehicle in a last step. In the following, those steps are explained in detail.

\subsubsection{Get Image}
The camera takes images and writes them in a queue. Those images are taken equidistant with a short exposure time. The time difference between the images is constant. 

\subsubsection{Find Points}
In a second step, we search for the LED points in the image. As the images are taken with short exposure times, the LEDs are detected robustly using OpenCV \citep{opencv_library}. We detect the contours of the LED points and determine their moments. The moments describe the centers of the blob. We filter the detected contours by size to remove disturbance points, i.e., blobs which are bigger or smaller than the expected size are dropped.

\subsubsection{Find Vehicles}
In a next step, we match points to vehicles. For this purpose, we use the known distances between the LEDs as a measure. We propose two methods.
First, we know the distance between the two LEDs on the back of the vehicle. In our case, the basis of the triangle. In the following, this distance is called vehicle-width. Furthermore, the longest distance between two LEDs on the same vehicle is known. This one is called vehicle-length. We use the vehicle-width to determine the vehicle back, i.e. the two LEDs mounted on the back of the vehicle. 

\begin{figure}
  \centering
  \input{tikz/BackMapping.tex}
  \caption{The green marked annulus is the area in the vehicle-width up to a tolerance to the green marked point. It includes points that may belong to the same vehicle back.}
  \label{Fig:BackMapping}
\end{figure}
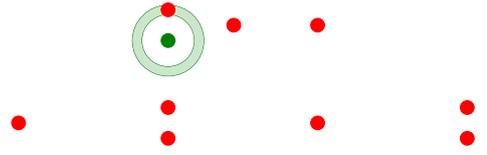
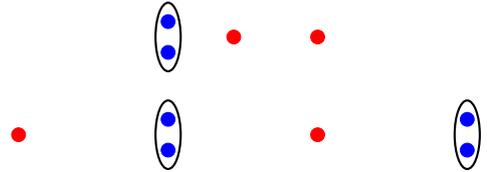
\begin{figure}
  \centering
  \input{tikz/Backs.tex}
  \caption{The black ellipses surround all vehicle backs found with the help of the vehicle-width. The blue points are the LEDs of the vehicle back.}
  \label{Fig:Backs}
\end{figure}

For each point, we compute all points exactly in its vehicle-width up to a tolerance. This is shown exemplary in Fig.~\ref{Fig:BackMapping}.
From the geometry of the LEDs on the vehicle, it holds that if a point has another point in its vehicle-width, they form a vehicle back. Hence, we can compute all vehicle backs as illustrated in Fig.~\ref{Fig:Backs}, i.e. the geometry avoids ambiguities in scenarios with multiple vehicles. 
Since the LEDs are mounted in a non-equilateral triangle, we receive the direction of the vehicle from the vehicle backs. With this knowledge, we can reduce the search space for the remaining points. Those points are orthogonal to the vehicle back in the vehicle-length. For this, we only consider those points including a tolerance, as shown in Fig.~\ref{Fig:Rectangles}. That means for all vehicle backs, we know with which other points they can form a vehicle, namely those inside the green shaped area in Fig.~\ref{Fig:Rectangles}.
\begin{figure}
  \centering
  \input{tikz/Rectangles.tex}
  \caption{The green cycles are the areas in which points belonging to the green vehicle back points are placed. They are in the vehicle-length orthogonal to the vehicle back, up to a tolerance.}
  \label{Fig:Rectangles}
\end{figure}
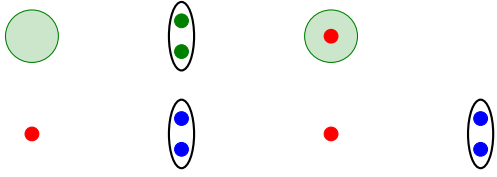
\begin{figure}
  \centering
  \input{tikz/VehicleMapping.tex}
  \caption{The green marked annulus includes the area in the vehicle-length from the green point. This is the area, in which vehicle backs are placed that may belong to the same vehicle.}
  \label{Fig:VehicleMapping}
\end{figure}
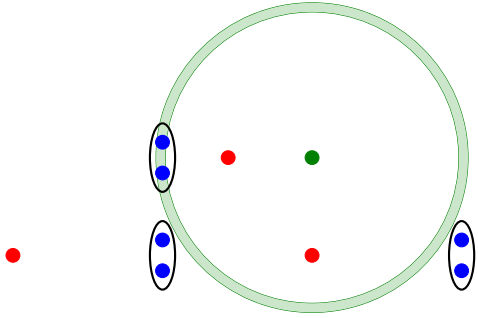
Now, we consider all points that do not belong to a vehicle back. For those, we have no restriction in direction to determine the points with which they can form a vehicle. Here, the only restriction is the vehicle-length. Hence, for each non-vehicle-back point, we consider all points in the vehicle-length as possible points with which it can form a vehicle. Those areas are shown for one point in Fig.~\ref{Fig:VehicleMapping}.
Now, we know for each point with which other points it can form a vehicle. Thus, if two points are not in the area of the other, they cannot be mounted on the same vehicle. That also means, if two LEDs are mounted on the same vehicle, the corresponding points are in the area of each other. We use this to determine the vehicles. For this purpose, we choose one point with the least possible matches. Then, we compute the intersection of its matches, including itself, with the matches of all points in its area. If this intersection is equivalent to the matches of the chosen point, the intersection only contains points belonging to the same vehicle. 


Since the points from the intersection are mapped to a vehicle, we remove all those points from the remaining areas. Then, the procedure of choosing a point and intersecting the areas is repeated until all points are mapped or no progress is reached. 

\begin{figure}
\centering
\input{tikz/Ambiguity.tex}
\caption{The annulus of the green point and the areas of the orange vehicle back points and the blue vehicle back points. The double-colored cycle is in the area of the orange vehicle back points and the blue vehicle back points. The second blue cycle is omitted due to space reasons.}
\label{fig:ambiguity}
\end{figure}
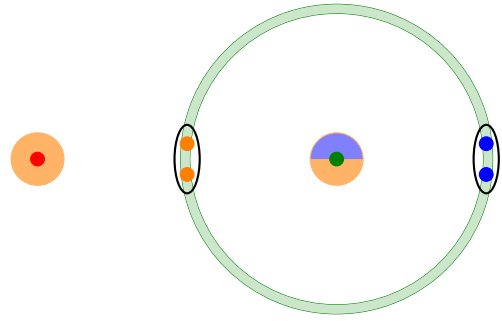

However, there may be points that are in each others area that are not mounted on the same vehicle, see Fig.~\ref{fig:ambiguity}. The green point has two vehicle backs in its area and is 
in the area of both vehicle backs, orange and blue. Therefore, the intersection of the sets of possible matches of the green point and the orange vehicle back is different from the original sets. Hence, they cannot be matched yet. Nevertheless, the intersection of the sets of the green point and the blue vehicle back is equal to the original set of the blue vehicle back. Thus, 
they may be matched. Once removed from the pool of unmatched points, the remaining points can be matched unambiguously.

For a small amount of vehicles it may be feasible to check every combination of points and check if its distances are feasible to match a vehicle. In this case, conflicts may occur where ambiguities occur in the previous procedure, e.g. in the example in Fig.~\ref{fig:ambiguity}. The check of combinations would detect three vehicles, the ones shown in the Fig.~\ref{fig:ambiguity} and a vehicle containing of the orange vehicle-back and the green front LED. Therefore, those conflicts have to be resolved until each LED point is part of only one vehicle. 

\subsubsection{Match Vehicles}
Now, we have found all vehicles in one image. To determine the frequency of the identification LED, we need the sequence of points for each vehicle. For this purpose, we need to match the vehicles in the current image to the corresponding vehicles in the previous image. We use that we receive the images with a high frequency. Hence, vehicles can only move a short distance between the images. We match each vehicle to the nearest vehicle in the previous image. To avoid false matching, some plausibility checks are done. For the position and the orientation, we check whether the observed change between the previous and the current image is physically possible. For identification, we check whether the recalculation of the identity yields the same result as before. 

\subsubsection{Compute Pose}
By now, we have the sequence of points for each vehicle in the different images. As a resulting step, we compute the pose of the vehicles. For this, we consider the three positioning points from the current image. If the identification point is visible in the latest image, it can be filtered by the sum of the distances of one point to all other. To compute the orientation, we consider the straight line between two LED points. Then, we can compute the angle between this line and the x-axis. Since we are interested in the orientation to the side of the vehicle, we determine the offset of the straight line to the side of the vehicle and add it to the previous computed angle. For more robustness, we compute the orientation for all three pairs of LEDs and take the median as result. In this way, we can remove outliers. The position of a vehicle is defined as its midpoint. Hence for positioning, we compute the midpoint of the two back points and shift this point to the midpoint of the vehicle using the previously calculated orientation.\\

\subsubsection{Identification}
\begin{figure}
\centering
\subfigure[Sampling when the LED is on or off.]{\includegraphics[width=0.65\columnwidth]{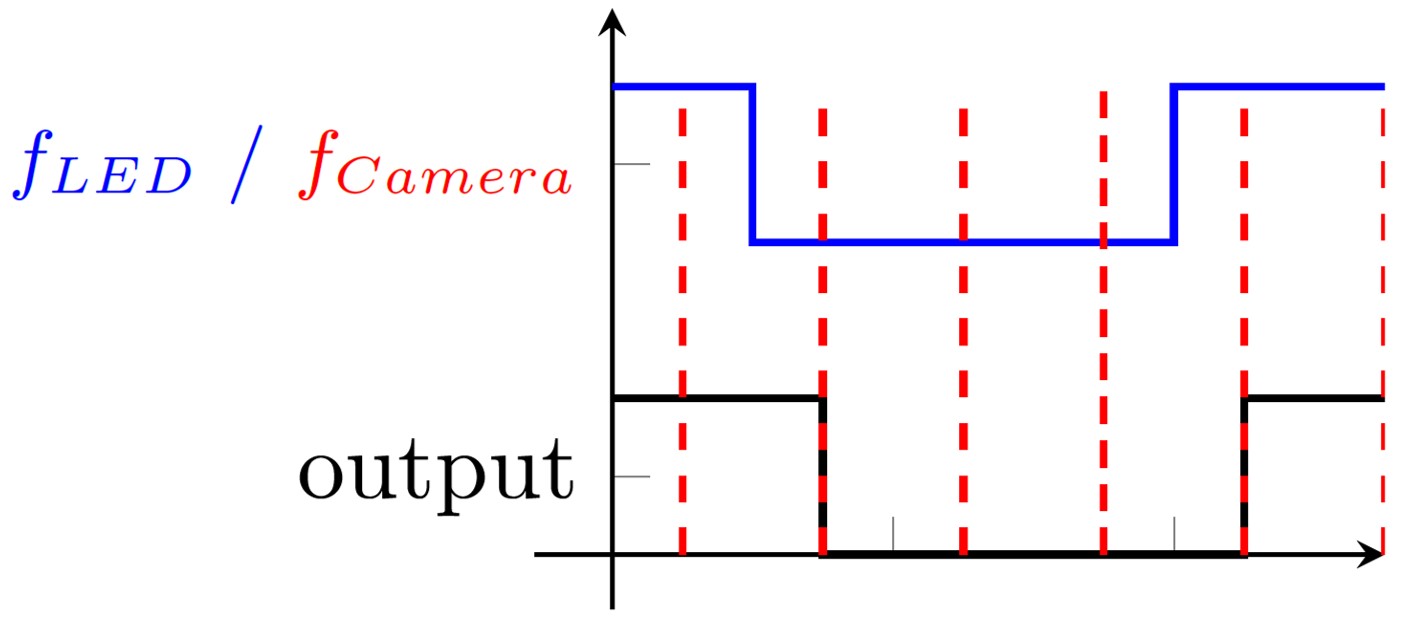} \label{fig:identity1}}
\subfigure[Sampling when the LED is turning on or off.]{\includegraphics[width=0.4\columnwidth]{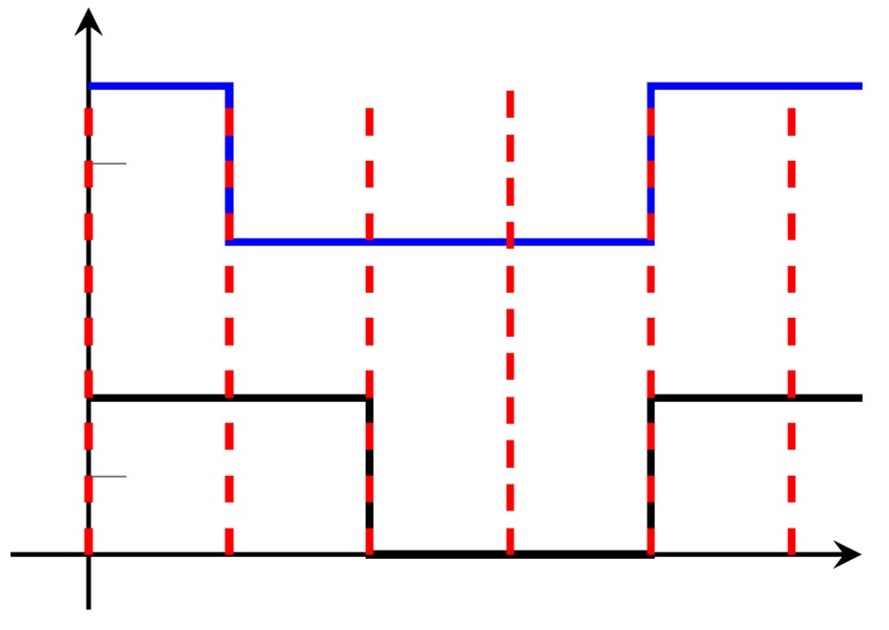} \label{fig:identity2}}
\subfigure[Sampling when the LED is turning on or off.]{\includegraphics[width=0.4\columnwidth]{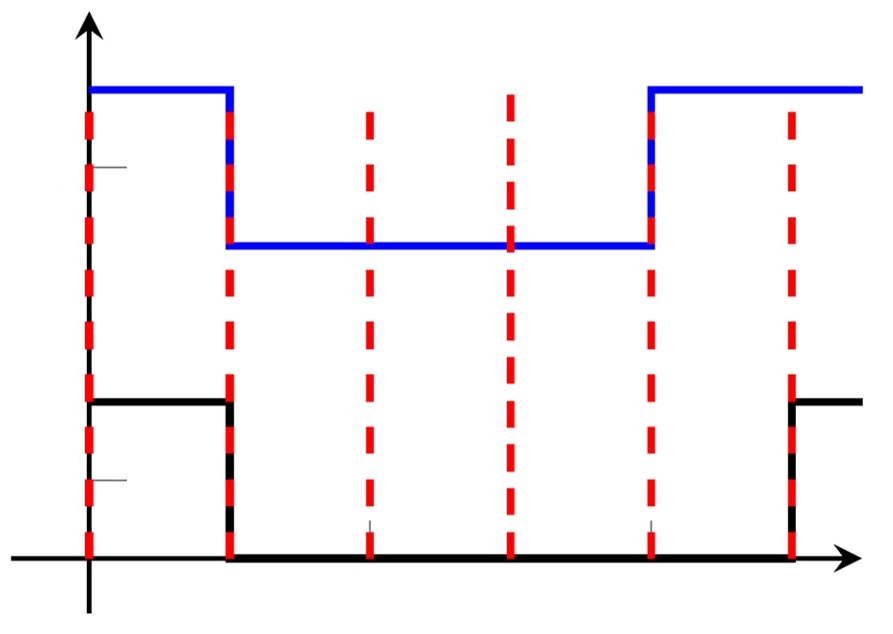} \label{fig:identity3}}
\caption{Possible situations while sampling. The dashed red lines depict the time points when an image with the camera is taken. The blue line describes the signal which is sampled and the black line the result of the sampling.}
\label{fig:identify}
\end{figure}

For the identification, we use the sequence of points. This sequence is received from the order of images received by the camera. For each vehicle, that sequence is an order of sets of three or four points each belonging to this vehicle in the respective image. To determine the identity, we count in how many consecutive images the identification LED is on $num_{on}$ and in how many it is off $num_{off}$. Using the camera frequency $f_{Camera}$, the time in which it is on or off $t_{on}/t_{off}$, respectively, can be computed by 
\begin{equation*}
t_{on/off} = \frac{num_{on/off}}{f_{Camera}} 
\end{equation*}
A mapping of LED frequencies to concrete IDs has to be provided to our system beforehand. For this approach, it is important that the LED frequencies have such high distances that they can be recognized uniquely. For this purpose, we choose the LED frequencies depending on the camera frequency and the number of images in which we expect the LED to be on or off. To guarantee that there is no overlapping, we only consider every third number for a sequence of images. For example, we expect that for the first ID it is on in two following up images and for the next ID that it is on in five following up images. With this, we can handle sampling during a switch of the LED value, as illustrated in Fig.~\ref{fig:identify}. In Fig.~\ref{fig:identity1}, the camera samples while the LEDs are on or off, while Fig.~\ref{fig:identity2} and \ref{fig:identity3} show scenarios where the camera samples while the LEDs are turning on or off. For robust recognition, we map the frequencies detected with the intermediate numbers to the nearest number. In the end, we have for each ID a number of images $n$ in which we expect it to be on. From this number we can compute the frequency of the LED with the camera frequency $f_{LED} =\frac{f_{Camera}}{n}$ and the interval of detected frequencies which we map to this ID $[\frac{f_{Camera}}{n+1}, \frac{f_{Camera}}{n-1}]$. Please note that we do not check the last images again. Instead, the number of images in which the identification LED is on and the overall amount of images are tracked during operation. This approach requires some initial time to initiate the IDs of the vehicles, i.e., the minimum number of images to distinguish all used IDs.

The algorithm is summarized in Algorithm~\ref{Alg:SortingStep} where $[v_1, \dots, v_n]$ describes a vehicle consisting of the image points $v_1, \dots, v_n$ belonging to its LED points, $g_1, \dots, g_n$  are the reference LED points as image points from the vehicle geometry, $dis$ computes all distances of the points in the provided set sorted in ascending order and $n_j$ is the size of the vehicle-mapping of $v_j$ with the corresponding points $p_{j1}, \dots, p_{jn_j-1}$.  

\input{algorithm.tex}

\section{Correctness} 
\label{sec:4}
\input{proof.tex}

\section{EVALUATION}
\label{sec:evaluation}
We evaluate our system in different scenarios in simulations as well as in experimental tests. In simulations, we simulate the camera and feed our system with black images with white points at the positions where the LED would be observed. This has the advantages over the real tests that we exactly know the ground truth. Hence, we have a perfect reference value. However, the real test provides more realistic results for the accuracy of our system.

\subsection{Scenarios} 
First, we have static scenarios. Here, we place the vehicle at nine different positions in the room. In each position, we place the vehicles in eight different orientations. We choose the angles $0, \pm \frac{\pi}{4}, \pm \frac{\pi}{4}, \pm \frac{3\pi}{4}$ and $\pi$. Furthermore, we evaluate dynamic scenarios. First, we force the vehicles drive along a straight line. Here, we test eight different lines in both directions, each with 0.5 m/s, 0.8 m/s and 1 m/s. In the experimental test, we additionally evaluate a vehicle driving along a circle with 0.5 m/s, 0.8 m/s, 1 m/s and 1.5 m/s in both directions. We evaluate a higher velocity in the circle experiments, as the distance is larger than in the straight line experiments. In the simulation, we test a vehicle driving an ellipse and afterwards an eight. Here, we simulate up to $20$ vehicles. In a first step, we simulate them standing in small clusters and in a second step driving the ellipse and the recumbent eight. Moreover, we evaluate scenarios with two vehicles simulation right-hand or left-hand traffic, two parallel lines or passing a parking vehicle in different angles. Overall, we simulate $175$ scenarios and test $128$ in experiments.

\subsection{Simulation Results} 
To create the images that are the input to our system, we compute the positions of each vehicle in the world depending on the path which they follow. Then, we calculate the LED positions from the vehicle positions. With the help of a camera calibration, those world point can be translated in the image plane. At those positions, we draw white blobs in a black image and provide it to the system. Hence, we only simulate the camera. For the identification LED, we compute whether the LED is on or off at a specific time point. Each image gets a time stamp.\\
To compute the accuracy of our system, we compare the expected position from which the LED positions are computed with the position received by our system. For this purpose, we compute the euclidean error. The results of all $175$ scenarios as described above are summarized in Table~\ref{Tab:AccSim}.
\begin{table}
\centering
\caption{Accuracy in the simulation for all 175 scenarios. There is a small error in the simulation results, since calibration is used for comparison to experiments.}
\input{tables/AccuracySimulation.tex}
\label{Tab:AccSim}
\end{table}

\subsection{Experimental Results}
\begin{figure}
\centering
\subfigure[Comparison of the positioning error. The error is given in \unit{cm}.]{\label{Fig:CompPos}\resizebox{0.8\linewidth}{!}{\input{tikz/StaticTestComparePosition.tex}}}
\subfigure[Comparison of the orientation error. The error is given in \unit{degree}.]{\label{Fig:CompOri}\resizebox{0.8\linewidth}{!}{\input{tikz/StaticTestCompareOrientation.tex}}}
\caption{Comparison of the accuracy results of the simulations to the results of the experiments for all 72 static scenarios.}
\label{Fig:CompStatic}
\end{figure}
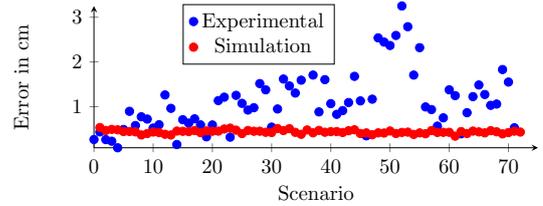
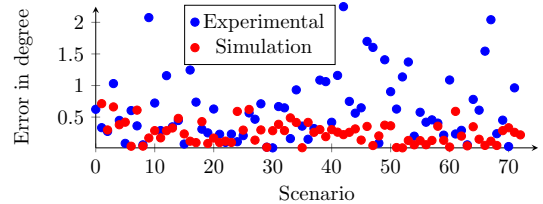
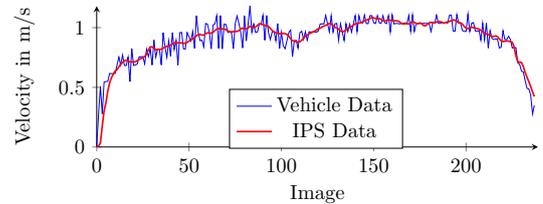
\begin{figure}
\centering
\resizebox{0.8\linewidth}{!}{\input{tikz/VelocityCompare.tex}}
\caption{Comparison between the velocity measured by the vehicle and the velocity detected by our system in the straight line experiment. Here, the distances of the image to its predecessor is illustrated}
\label{Fig:VelComp}
\end{figure}
Beside the simulation, we evaluate different scenarios in experimental tests with a model-scale vehicle and a camera in a height of about \unit[3]{m}. For the static scenarios, we place a vehicle at specific poses in the room. Then, we compare the pose gained by our system with the one that we intended to place. In contrast to the simulation, we cannot guarantee that the vehicle is placed exactly in the intended pose. Hence, an error may be introduced. In Fig.~\ref{Fig:CompStatic}, the computations in the simulations are compared to the results in the experiments for all $72$ static scenarios. \\
In the dynamic scenarios the placement in reality gets even worse. Beside the inaccurate placement, we need to know the poses at the different time points depending on the velocity. For this, we use data from odometer and \gls{IMU} on the vehicle to compute the reference value. To force the vehicle driving a straight line, we use a rail. For the circle, we fix a midpoint and attach a cord to this midpoint as well as to the vehicle. If the vehicle drives straight ahead, it is pulled along a circle. In Table~\ref{Tab:AccReal}, the accuracy results for the static test as well as for the lines and the circles are shown. Fig.~\ref{Fig:VelComp} compares the speed received from the odometer to the velocity computed from our \gls{IPS} for the straight line experiment. Here, we can see that those are comparable. However, there are some differences, e.g. at the start of the experiment. The tires start spinning, but the vehicle is not moving. The speed measured by the odometer rises, but the \gls{IPS} does not measure any movement. 
\begin{table}
\centering
\caption{Accuracy of the dynamic scenarios evaluating experimental results compared to the results of the static scenarios evaluated experimentally.}
\input{tables/AccuracyReality.tex}
\label{Tab:AccReal}
\end{table}
The errors for the static tests and the straight lines are comparable. However, the errors of the circles are worse. This is because the placement of the circles was the most difficult. If the length of the cord is measured only a few millimeters too short or too long, the error of the intended position to the actual position increases with covered distance. Furthermore since the vehicle drives straight ahead, the actual angle of the vehicle does not fit to the angle of the tangent to the circle. However for the straight lines, the results are comparable to the static tests. Here, we only have a single computation yielding a wrong result. This, we consider as an outlier. For all other computations, the maximal errors are in the range of the maximal errors of the static experiments.\\
Overall, the errors of the experimental test increase compared to the simulation. On the one hand, this is because the placement is more difficult as described above. On the other hand, the error of the calibration of the camera introduces an error in the positioning. Since no mapping between the world points of the camera and the world coordinate system used by the system is done in the simulation, there is no such calibration error.

\subsection{Efficiency}
\begin{figure}
\centering
\resizebox{.55\linewidth}{!}{\input{tikz/EfficiencyPerStep.tex}}
\caption{Mean latencies of the single steps of the algorithm for 3, 16 and 20 vehicles in the average case and in the worst case.}
\label{Fig:EffPerStep}
\end{figure}
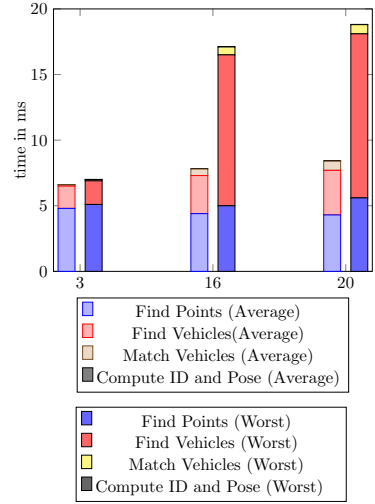

The next property of our system that we evaluate is the efficiency. For this purpose, we determine the average and the worst case. In the worst case, we have all vehicles driving in a platoon. Then, we can determine in the 'Find Vehicle'\textminus step only two vehicles at the same time namely the head and the tail of the platoon. Hence, we need many iterations to detect all vehicles. In contrast to this, the vehicles are clustered in small groups in the average case. In Fig.~\ref{Fig:EffPerStep}, the computations times for the different steps are shown for different number of vehicles in the worst case as well as in the average case. We can see that finding points in the image and computing the ID and the pose is constant for increasing number of vehicles as well as for the average and the worst case. Matching the vehicles is constant for the average case compared to the worst case, but increasing with the number of vehicles. This is because we need to match more vehicles. However, it has not influenced how those vehicles are positioned. The step with the most influence on the overall runtime is finding the vehicles. With increasing number of vehicles, the computation time increases. Furthermore in the worst case, the computations are higher than in the average case. This is because more vehicles need to be found with increasing number of vehicles and it is more difficult to find them in the worst case compared to the average case. 
Furthermore, we compare the runtimes with increasing number of vehicles for the average case and the worst case. The results can be seen in Fig.~\ref{Fig:EffAvCase} and Fig.~\ref{Fig:EffWoCase}, respectively. In the average case, all computations terminate within the soft deadline of \unit[20]{ms}. In the worst case, all computations for less than $14$ vehicles also terminate within the deadline. For more than $14$ vehicles, the deadline is exceeded. However, for $20$ vehicles \unit[87.57]{\%} of the computations terminate within the deadline. After computation, there is a delay of about 5 ms for communications of the poses to the vehicles. 
\begin{figure}
\centering
\resizebox{.4\linewidth}{!}{\input{tikz/EfficiencyAverageCase.tex}}
\caption{Mean and max latencies of our algorithm in the average case (vehicles split in clusters) for different number of vehicles and clusters. The deadline gained from the vehicle cycle time is marked as dashed red line.}
\label{Fig:EffAvCase}
\end{figure}
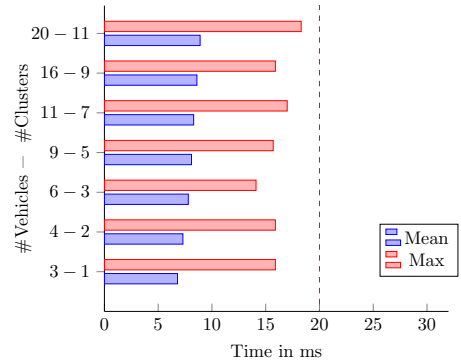
\begin{figure}
\centering
\resizebox{.5\linewidth}{!}{\input{tikz/EfficiencyWorstCase.tex}}
\caption{Mean and max latencies of our algorithm in the worst case (driving in a platoon) for different number of vehicles. The deadline gained from the vehicle cycle time is marked as dashed red line. The percentages describe the amount of computations that terminate within the 20 ms.}
\label{Fig:EffWoCase}
\end{figure}
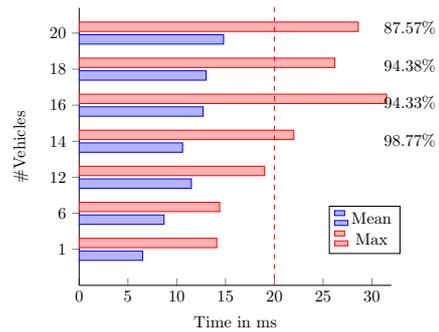

\subsection{Robustness}
\begin{figure}
\centering
\includegraphics[width=0.5\columnwidth]{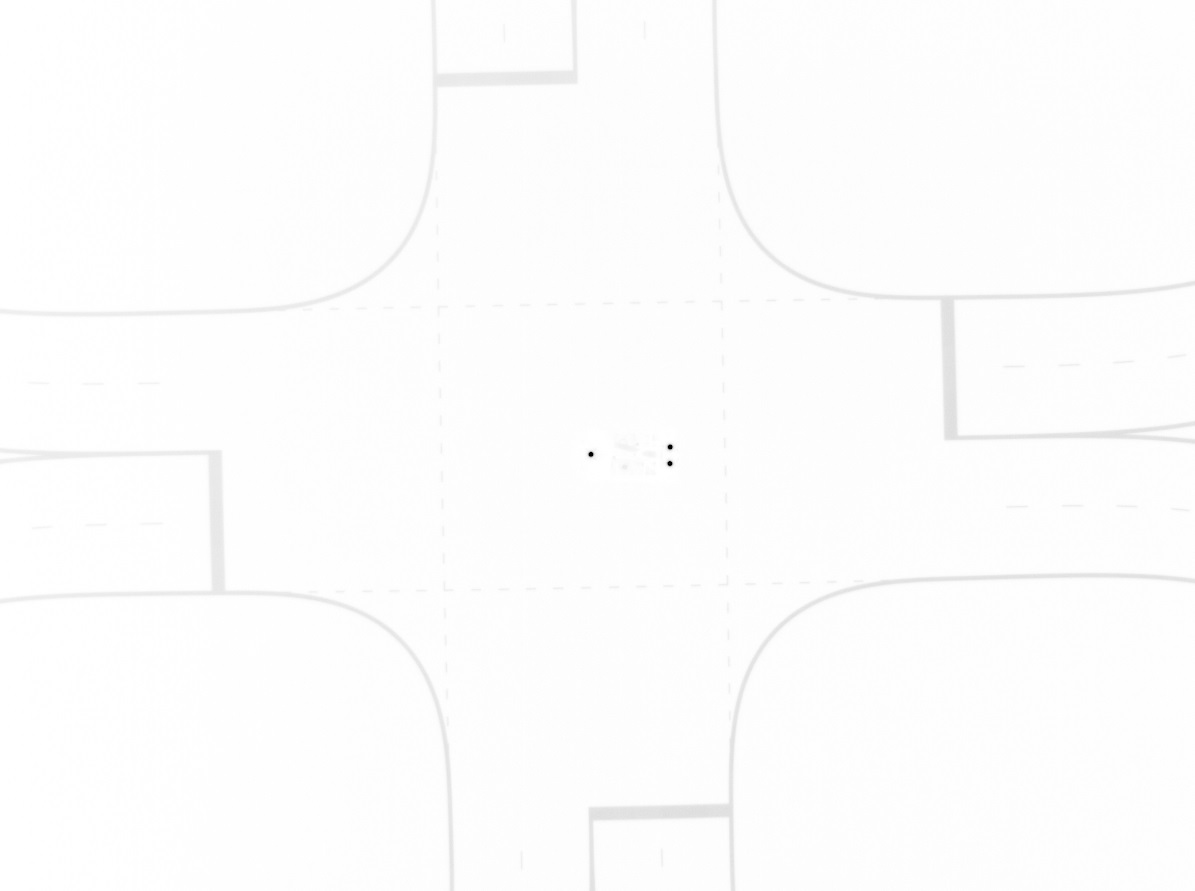}
\caption{A model-scale vehicle on an intersection, as seen from the \gls{IPS}. The colors are inverted due to print reasons.}
\label{fig:robust}
\end{figure}

We also evaluated the robustness of our system against changes of the lighting conditions. Due to the low exposure time and the light LEDs, the vehicles are detected robustly even in light environments. Fig.~\ref{fig:robust} shows an image of a vehicle on an intersection as seen from the \gls{IPS}. Please note that the colors are inverted. The white lines of the road are light grey, while the LED points of the vehicle are black. Hence, we can easily filter LED points of the vehicles using a threshold. 

However, if an led is broken or occluded, it can not be detected. For robustness aginst such errors, we implemented a timeout on the vehicles. If the \gls{IPS} is not able to detect a vehicle, the points are omitted, as stated in Section~\ref{sec:3}. Then, the vehicle does not receive any pose update for some time. If the vehicle does not receive an update for 100 ms, i.e. of 5 consequtive images, the vehicle stops. If the identification LED is occluded, the \gls{IPS} may detect a vehicle with a wrong ID. Therefore, one vehicle will not receive pose updates and will stop. 

\section{CONCLUSION}
\label{sec:conclusion}
We developed a new indoor positioning system which externally computes the position and the orientation of multiple vehicles. We evaluated our system with $20$ vehicles. Our system is real-time capable with a soft deadline of \unit[20]{ms}. Moreover, we reach an accuracy of around \unit[1]{cm} and of about \unit[0.6]{\degree} in the mean. We robustly detect all vehicles in the plane even with changing lighting conditions and reflections. While evaluating our system, no identification errors occurred. Hence, our indoor positioning system can be used in applications with model-scale autonomous vehicles in which the knowledge of the position of each vehicle is crucial. 

\normalem 
\bibliography{root}             
                                                   
\end{document}

%% file: tikz/BackMapping.tex
\begin{tikzpicture} [scale=0.12]

	\path[use as bounding box] (-25,6) rectangle (25, -15);

	\coordinate (v1) at (-8.2, 1.7);
	\coordinate (v2) at (-8.2, -1.7);
	\coordinate	(v3) at (8.2, 0);
	\coordinate (vI) at (-1, 0);
	
	\coordinate (v4) at (-8.2,-9.1);
	\coordinate (v5) at (-8.2,-12.5);
	\coordinate (v6) at (-24.6, -10.8);
	
	\coordinate	(v7) at (8.2, -10.8);
	\coordinate (v8) at (24.6,-9.1);
	\coordinate (v9) at (24.6,-12.5);
	
	\draw[darkgreen] (v2) circle (2.9);
	\draw[darkgreen] (v2) circle (3.9);
		
	\fill[darkgreen!20,even odd rule] (v2) circle (2.9) (v2) circle (3.9);
	
	\draw[red,fill=red] (v1) circle (5ex);
	\draw[darkgreen,fill=darkgreen] (v2) circle (5ex);
	\draw[red,fill=red] (v3) circle (5ex);
	\draw[red,fill=red] (vI) circle (5ex);
	\draw[red,fill=red] (v4) circle (5ex);
	\draw[red,fill=red] (v5) circle (5ex);
	\draw[red,fill=red] (v6) circle (5ex);
	\draw[red,fill=red] (v7) circle (5ex);
	\draw[red,fill=red] (v8) circle (5ex);
	\draw[red,fill=red] (v9) circle (5ex);

\end{tikzpicture}

%% file: tikz/Backs.tex
\begin{tikzpicture} [scale=0.12]

	\path[use as bounding box] (-25,6) rectangle (25, -15);

	\coordinate (v1) at (-8.2, 1.7);
	\coordinate (v2) at (-8.2, -1.7);
	\coordinate	(v3) at (8.2, 0);
	\coordinate (vI) at (-1, 0);
	
	\coordinate (v4) at (-8.2,-9.1);
	\coordinate (v5) at (-8.2,-12.5);
	\coordinate (v6) at (-24.6, -10.8);
	
	\coordinate	(v7) at (8.2, -10.8);
	\coordinate (v8) at (24.6,-9.1);
	\coordinate (v9) at (24.6,-12.5);

	\draw[blue,fill=blue] (v1) circle (5ex);
	\draw[blue,fill=blue] (v2) circle (5ex);
	\draw[red,fill=red] (v3) circle (5ex);
	\draw[red,fill=red] (vI) circle (5ex);
	\draw[blue,fill=blue] (v4) circle (5ex);
	\draw[blue,fill=blue] (v5) circle (5ex);
	\draw[red,fill=red] (v6) circle (5ex);
	\draw[red,fill=red] (v7) circle (5ex);
	\draw[blue,fill=blue] (v8) circle (5ex);
	\draw[blue,fill=blue] (v9) circle (5ex);
	
	\node[fit=(v1)(v2),ellipse, draw=black, thick] {};
	\node[fit=(v4)(v5),ellipse, draw=black, thick] {};
	\node[fit=(v8)(v9),ellipse, draw=black, thick] {};

\end{tikzpicture}

%% file: tikz/Rectangles.tex
\begin{tikzpicture} [scale=0.12]

	\path[use as bounding box] (-25,6) rectangle (25, -15);

	\coordinate (v1) at (-8.2, 1.7);
	\coordinate (v2) at (-8.2, -1.7);
	\coordinate	(v3) at (8.2, 0);
	\coordinate (vI) at (-1, 0);
	
	\coordinate (v4) at (-8.2,-9.1);
	\coordinate (v5) at (-8.2,-12.5);
	\coordinate (v6) at (-24.6, -10.8);
	
	\coordinate	(v7) at (8.2, -10.8);
	\coordinate (v8) at (24.6,-9.1);
	\coordinate (v9) at (24.6,-12.5);


	\node[fit=(v1)(v2),ellipse, draw=black, thick, fill=white] {};
	\node[fit=(v4)(v5),ellipse, draw=black, thick, fill=white] {};
	\node[fit=(v8)(v9),ellipse, draw=black, thick, fill=white] {};

	\fill[darkgreen!20] (-24.6, 0) circle (2.9);
	\fill[darkgreen!20] (v3) circle (2.9);
	\draw[darkgreen] (-24.6, 0) circle (2.9);
	\draw[darkgreen] (v3) circle (2.9);

	\draw[darkgreen,fill=darkgreen] (v1) circle (5ex);
	\draw[darkgreen,fill=darkgreen] (v2) circle (5ex);
	\draw[red,fill=red] (v3) circle (5ex);
	\draw[blue,fill=blue] (v4) circle (5ex);
	\draw[blue,fill=blue] (v5) circle (5ex);
	\draw[red,fill=red] (v6) circle (5ex);
	\draw[red,fill=red] (v7) circle (5ex);
	\draw[blue,fill=blue] (v8) circle (5ex);
	\draw[blue,fill=blue] (v9) circle (5ex);
	
\end{tikzpicture}

%% file: tikz/VehicleMapping.tex
\begin{tikzpicture} [scale=0.12]

	\path[use as bounding box] (-25,17) rectangle (25, -18);

	\coordinate (v1) at (-8.2, 1.7);
	\coordinate (v2) at (-8.2, -1.7);
	\coordinate	(v3) at (8.2, 0);
	\coordinate (vI) at (-1, 0);
	
	\coordinate (v4) at (-8.2,-9.1);
	\coordinate (v5) at (-8.2,-12.5);
	\coordinate (v6) at (-24.6, -10.8);
	
	\coordinate	(v7) at (8.2, -10.8);
	\coordinate (v8) at (24.6,-9.1);
	\coordinate (v9) at (24.6,-12.5);

	\draw[darkgreen] (v3) circle (17.1);
	\draw[darkgreen] (v3) circle (16.1);
	\fill[darkgreen!20,even odd rule] (v3) circle (16.1) (v3) circle (17.1);
	
	\draw[blue,fill=blue] (v1) circle (5ex);
	\draw[darkgreen,fill=darkgreen] (v3) circle (5ex);
	\draw[red,fill=red] (vI) circle (5ex);
	\draw[blue,fill=blue] (v2) circle (5ex);
	\draw[blue,fill=blue] (v4) circle (5ex);
	\draw[blue,fill=blue] (v5) circle (5ex);
	\draw[red,fill=red] (v6) circle (5ex);
	\draw[red,fill=red] (v7) circle (5ex);
	\draw[blue,fill=blue] (v8) circle (5ex);
	\draw[blue,fill=blue] (v9) circle (5ex);
		
	\node[fit=(v1)(v2),ellipse, draw=black, thick] {};
	\node[fit=(v4)(v5),ellipse, draw=black, thick] {};
	\node[fit=(v8)(v9),ellipse, draw=black, thick] {};	

\end{tikzpicture}

%% file: tikz/Ambiguity.tex
\begin{tikzpicture} [scale=0.12]

	\path[use as bounding box] (-25,4.8) rectangle (25, -29);

	\coordinate (v1) at (-8.2, 1.7);
	\coordinate (v2) at (-8.2, -1.7);
	\coordinate	(v3) at (8.2, 0);
	\coordinate (vI) at (-1, 0);
	
	\coordinate (v4) at (-8.2,-9.1);
	\coordinate (v5) at (-8.2,-12.5);
	\coordinate (v6) at (-24.6, -10.8);
	
	\coordinate	(v7) at (8.2, -10.8);
	\coordinate (v8) at (24.6,-9.1);
	\coordinate (v9) at (24.6,-12.5);

	\draw[darkgreen] (v7) circle (17.1);
	\draw[darkgreen] (v7) circle (16.1);
	\fill[darkgreen!20,even odd rule] (v7) circle (16.1) (v7) circle (17.1);
	
	\fill[orange!60] (v6) circle (2.9);
	\fill[orange!60] (v7) circle (2.9);
	\fill[blue!50] (v7) -- (5.3,-10.8) arc (180:0:2.9) -- (v7);
	\draw[orange!60] (v6) circle (2.9);
	\draw[orange!60] (v7) circle (2.9);

	\draw[darkgreen,fill=darkgreen] (v7) circle (5ex);
	\draw[orange,fill=orange] (v4) circle (5ex);
	\draw[orange,fill=orange] (v5) circle (5ex);
	\draw[red,fill=red] (v6) circle (5ex);
	\draw[blue,fill=blue] (v8) circle (5ex);
	\draw[blue,fill=blue] (v9) circle (5ex);
		
	\node[fit=(v4)(v5),ellipse, draw=black, thick] {};
	\node[fit=(v8)(v9),ellipse, draw=black, thick] {};

\end{tikzpicture}

%% file: algorithm.tex
\normalem
\IncMargin{2em}
\begin{algorithm}[t]
	\SetKwInOut{Input}{Input}
    \SetKwInOut{Output}{Output}

\Indm
 \Input{Vehicle Mapping for each point and a point $p$}
 \Output{Whether $p$ belongs to a vehicle, is a disturbance point or cannot be mapped, yet. If $p$ belongs to a vehicle, the corresponding vehicle.}
\Indp
 get $\lbrace [p], p_1, \dots, p_{n-1} \rbrace$\;
 \lIf {$n < 3$} {\Return disturbance point}
 \lIf {$n > 4$} {\Return cannot be mapped yet}
 $S = \bigcap\limits_{i=1}^{n-1} \lbrace [p_i],p_{i1}, \dots, p_{im_i} \rbrace \cap \lbrace [p], p_1, \dots, p_{n-1} \rbrace$\;
\If {$S == \lbrace [p], p_1, \dots, p_{n-1} \rbrace$} {
	\lIf {$dis(S) \approx dis(\lbrace g_1, \dots, g_n \rbrace)$} { \Return Vehicle $[p,  p_1, \dots, p_{n-1}]$}
	\lIf {$n == 3$} { \Return disturbance point}
	\Return cannot be mapped yet\;
}
\lIf {$n == 3$} { \Return disturbance point}
\Return cannot be mapped yet\;
 \caption[Sorting step]{One step of matching a set of points to a vehicle.}
 \label{Alg:SortingStep}
\end{algorithm}
\ULforem
\DecMargin{2em}

%% file: proof.tex
In the following, we prove that this computation is correct. That means, we prove the following theorem.

\begin{thm}
\label{Theorem}
Assume that disturbance points do not form exactly a vehicle geometry with other points. Then, it holds that:\\
If our system detects a vehicle consisting of the points $v_1, \dots, v_n$ with $n=3$ or $n=4$, the LEDs corresponding to these points are mounted on the same vehicle.
\end{thm}

First, we formalize some properties. From the reality, we know that if we have a vehicle, all LEDs on the vehicle are at most in the vehicle-length to each other and the distance of the LED points match the distances of the vehicle geometry. The assumption yields the other direction. We assume that if a set of points is in the vehicle-length of each other and all the distances match the vehicle geometry, the set of points belongs to one vehicle. Formally,
\begin{equation} \label{eq:assumption}
\begin{split}
[v_1, \dots, v_n] \Leftrightarrow &\forall i=1\dots n \ldotp \forall j=1\dots n, j \neq i\ldotp \\ &\hspace{0.8cm}v_i \in \lbrace [v_j], p_{j1}, \dots, p_{jn_j-1} \rbrace\\ &\wedge dis(\lbrace  v_1, \dots, v_n \rbrace) = dis(\lbrace  g_1, \dots, g_n \rbrace)
\end{split}
\end{equation}

For proving the Theorem \ref{Theorem}, we first show the following lemma.

\begin{lemma}
\label{Lemma}
Assume Equation (\ref{eq:assumption}). Then, it holds that:\\
If Algorithm \ref{Alg:SortingStep} forms a vehicle $[p, v_1, \dots, v_{n-1}]$ with $n=3$ or $n=4$ for a provided point $p$, the LEDs corresponding to these points are mounted on the same vehicle.
\end{lemma}

\begin{pf}
We prove that if the algorithm matches a point to a vehicle, the found vehicle is indeed a vehicle by a case distinction on $n$. So let $p$ be an arbitrary point with
\begin{equation} \label{eq:point_p}
\lbrace [p], p_1, \dots, p_{n-1} \rbrace.
\end{equation}
\begin{case}
Let $n < 3$.\\
$p$ is filtered out as disturbance point by line $2$ of Algorithm \ref{Alg:SortingStep}. Thus, no vehicle is detected and especially no wrong vehicle.
\end{case}

\begin{case} 
Let $ n= 3$ with $\lbrace [p],q, r \rbrace$.
\begin{itemize}
\item[{a)}] Let the LEDs corresponding to $p,q,r$ be mounted on the same vehicle\\
Thus, we know from Equation (\ref{eq:assumption}) that 
\begin{align*}
p &\in \lbrace [r],r_1, \dots, r_{m-1} \rbrace\\
p &\in \lbrace [q],q_1, \dots, q_{k-1} \rbrace \\
q &\in \lbrace [r],r_1, \dots, r_{m-1} \rbrace \\
r &\in \lbrace [q],q_1, \dots, q_{k-1} \rbrace
\end{align*}
with this and Equation (\ref{eq:point_p}) we have 
\begin{align} \label{eq:intersec1}
\lbrace p, q, r \rbrace &\subseteq \lbrace [r],r_1, \dots, r_{m-1} \rbrace \cap  \lbrace [q],q_1, \dots, q_{k-1} \rbrace .
\end{align}
Furthermore, the set $S$ from Algorithm \ref{Alg:SortingStep} line $4$ intersecting the above intersection with the mapping of $p$ is
\begin{align*}
S = \lbrace [r],r_1, \dots, r_{m-1} \rbrace \cap  \lbrace [q],q_1, \dots, q_{k-1} \rbrace \cap \lbrace [p],q, r \rbrace .
\end{align*}
Thus, from Equation (\ref{eq:intersec1}) we receive that
\begin{align*}
S =  \lbrace [p],q, r \rbrace .
\end{align*}
Now, the algorithm checks in line 6 the distances. Thus, the distances of the points $dis(\lbrace p, q, r \rbrace)$ are compared to the distances of the position points $dis(\lbrace g_1, g_2, g_3 \rbrace)$. Since the LEDs corresponding to the points $p,q,r$ are mounted on the same vehicle, we know from Equation (\ref{eq:assumption}) that the distances are similar. Therefore, the Algorithm \ref{Alg:SortingStep} finds the vehicle $[p,q,r]$ which is indeed a vehicle.
\item[{b)}] Let the light points corresponding to the points $p,q,r$ do not be mounted on the same vehicle.\\
\begin{itemize}
\item[{i)}] $\exists s\in \lbrace p,q,r \rbrace \ldotp \exists t\in \lbrace p,q,r\rbrace\backslash \lbrace s \rbrace \ldotp$\\ $s \not\in \lbrace [t], t_1, \dots, t_{l-1} \rbrace$\\
Thus, the intersection does not contain one of the points $p,q,r$ that means
\begin{align*}
\lbrace p, q, r  \rbrace &\not\subseteq S \\ &= \lbrace [r],r_1, \dots, r_{m-1} \rbrace \cap  \lbrace [q],q_1, \dots, q_{k-1} \rbrace\\ &\cap \lbrace [p],q, r \rbrace
\end{align*}
Therefore, $ S \neq \lbrace [p],q, r \rbrace$. In that case, the Algorithm \ref{Alg:SortingStep} determines $p$ as a disturbance point in line $10$ and does not detect a wrong vehicle.
\item[{ii)}] $\forall s\in \lbrace p,q,r \rbrace \ldotp \forall t\in \lbrace p,q,r\rbrace\backslash \lbrace s \rbrace \ldotp \\ s \in \lbrace [t], t_1, \dots, t_{l-1} \rbrace$\\
From Equation (\ref{eq:assumption}), we then know that
\begin{align*}
dis(\lbrace [r],r_1, \dots, r_{m-1} \rbrace &\cap  \lbrace [q],q_1, \dots, q_{k-1} \rbrace\\ &\cap \lbrace [p],q, r \rbrace)\\ &\neq dis(\lbrace g_1, g_2, g_3 \rbrace) .
\end{align*}
Thus, the Algorithm \ref{Alg:SortingStep} detects $p$ as disturbance point in line $7$ and does not detect a wrong vehicle.
\end{itemize}
\end{itemize}
\end{case}

\begin{case}
Let $ n = 4$ with $\lbrace [p], q, r, s \rbrace$.
\begin{itemize}
\item[a)] Let the LEDs corresponding to $p,q,r,s$ be mounted on the same vehicle\\
This case is equivalent to {Case 2a)}. The set $S$ is equal to $\lbrace [p], q, r, s \rbrace$ and since the points $p,q,r,s$ belong to a vehicle, the distances are comparable to $dis(\lbrace g_1, g_2, g_3, g_4 \rbrace)$. Thus, the algorithm finds the vehicle $[p,q,r,s]$ which is indeed a vehicle.
\item[b)] Let the light points corresponding to the points $p,q,r,s$ do not be mounted on the same vehicle.\\
Analog to {Case 2b)}, we need to distinguish two further cases.
\begin{itemize}
\item[i)] $\exists u\in \lbrace p,q,r,s \rbrace \ldotp \exists t\in \lbrace p,q,r,s\rbrace\backslash \lbrace u \rbrace \ldotp$ \\ $u \not\in \lbrace [t], t_1, \dots, t_{l-1} \rbrace$\\
Thus analog to {Case 2b)i)}, the set $S \neq \lbrace [p], q, r, s \rbrace$. Therefore, the Algorithm \ref{Alg:SortingStep} cannot map the point $p$ yet and terminates in line $11$ without finding a wrong vehicle.
\item[ii)]$\forall s\in \lbrace p,q,r \rbrace \ldotp \forall t\in \lbrace p,q,r\rbrace\backslash \lbrace s \rbrace \ldotp$\\ $s \in \lbrace [t], t_1, \dots, t_{l-1} \rbrace$\\
Thus with the same argumentation as in {Case 2b)ii)}, the algorithm cannot map $p$ yet and terminates in line $8$ without finding a wrong vehicle.
\end{itemize}
\end{itemize}
\end{case}

\begin{case}
Let $ n > 4$\\
$p$ cannot be mapped yet by Algorithm \ref{Alg:SortingStep} and thus it terminates in line $3$ without finding a wrong vehicle.
\end{case}

Thus, in all cases if a vehicle was found, it was indeed a vehicle. 
\end{pf}

From Lemma \ref{Lemma}, we can conclude that the Theorem \ref{Theorem} holds. As the lemma shows that for one point $p$ only a correct vehicle is found by Algorithm \ref{Alg:SortingStep}, we can extend this to all points since $p$ was an arbitrary point and Algorithm \ref{Alg:SortingStep} is the only part of our system that recognizes vehicles.\\
Thus, we have shown that if a vehicle is found by our algorithm, these points indeed form a vehicle.

%% file: tables/AccuracySimulation.tex
\begin{tabular}{ l | l l}
			& Position  & Orientation\\
			& Error in \unit{cm} &  Error in \unit{\degree}\\
			\hline
 	Mean  & $0.449673$ & $0.256081$\\
 	Max   & $0.569379$ 		& $0.876293$\\
 	Std dev  & $0.0405785$ 	& $0.153931$\\
\end{tabular}

%% file: tikz/StaticTestComparePosition.tex
\begin{tikzpicture}
\begin{axis}  [
width = 0.5\textwidth,
height=4cm,
xlabel={Scenario},
ylabel={Error in \unit{cm}},
xlabel style={below},
axis x line=left,
axis y line=left,
xmax=75,
y label style={at={(0,0.5)}},
legend style={at={(0.2,0.8)},anchor=west}
]
\addplot[only marks, color=blue] coordinates {
(0,0.264662)
(1,0.434594)
(2,0.26233)
(3,0.226635)
(4,0.0798731)
(5,0.489495)
(6,0.895327)
(7,0.57671)
(8,0.773144)
(9,0.721549)
(10,0.520634)
(11,0.595484)
(12,1.26165)
(13,0.961781)
(14,0.152726)
(15,0.708059)
(16,0.631676)
(17,0.727671)
(18,0.591821)
(19,0.324696)
(20,0.590388)
(21,1.13514)
(22,1.21182)
(23,0.317424)
(24,1.2528)
(25,1.07386)
(26,0.923448)
(27,0.976119)
(28,1.51558)
(29,1.37616)
(30,0.54204)
(31,0.947774)
(32,1.61719)
(33,1.46339)
(34,1.30411)
(35,1.59186)
(36,2.06163)
(37,1.70649)
(38,0.882587)
(39,1.60368)
(40,1.06667)
(41,0.828235)
(42,0.91129)
(43,1.09178)
(44,1.6774)
(45,1.12886)
(46,0.34677)
(47,1.1668)
(48,2.53792)
(49,2.44375)
(50,2.36746)
(51,2.59223)
(52,3.25062)
(53,2.78774)
(54,1.70577)
(55,2.31871)
(56,0.993753)
(57,0.93369)
(58,0.5641)
(59,0.749791)
(60,1.37649)
(61,1.24669)
(62,0.392174)
(63,0.864141)
(64,1.22372)
(65,1.48465)
(66,1.27121)
(67,1.02915)
(68,1.05764)
(69,1.82962)
(70,1.54978)
(71,0.527001)
	};
	
\addplot[only marks, color=red] coordinates {
(1,0.532995)
(2,0.466779)
(3,0.491497)
(4,0.481398)
(5,0.442453)
(6,0.444392)
(7,0.435614)
(8,0.366385)
(9,0.397939)
(10,0.431696)
(11,0.41962)
(12,0.382088)
(13,0.365674)
(14,0.453695)
(15,0.448159)
(16,0.441942)
(17,0.465482)
(18,0.418204)
(19,0.45624)
(20,0.44998)
(21,0.452173)
(22,0.49962)
(23,0.518817)
(24,0.473965)
(25,0.397284)
(26,0.467298)
(27,0.449455)
(28,0.450256)
(29,0.426299)
(30,0.420799)
(31,0.510597)
(32,0.465025)
(33,0.507172)
(34,0.425193)
(35,0.380949)
(36,0.478916)
(37,0.414009)
(38,0.468298)
(39,0.427387)
(40,0.440591)
(41,0.442584)
(42,0.421187)
(43,0.461173)
(44,0.4823)
(45,0.398921)
(46,0.412578)
(47,0.370423)
(48,0.413591)
(49,0.409564)
(50,0.455348)
(51,0.398714)
(52,0.423643)
(53,0.429723)
(54,0.378436)
(55,0.412627)
(56,0.393658)
(57,0.464179)
(58,0.422662)
(59,0.425333)
(60,0.426192)
(61,0.339563)
(62,0.441921)
(63,0.398447)
(64,0.452638)
(65,0.434704)
(66,0.413223)
(67,0.462176)
(68,0.437521)
(69,0.390583)
(70,0.419653)
(71,0.451625)
(72,0.43105)
	};
	
\legend{Experimental, Simulation}

\end{axis}
\end{tikzpicture}

%% file: tikz/StaticTestCompareOrientation.tex
\begin{tikzpicture}
\begin{axis} [
width = 0.5\textwidth,
height=4cm,
xlabel={Scenario},
ylabel={Error in degree},
xlabel style={below},
axis x line=left,
axis y line=left,
xmax=75,
y label style={at={(0,0.5)}},
legend style={at={(0.2,0.8)},anchor=west}
]
\addplot[only marks, color=blue] coordinates {
(0,0.623741)
(1,0.332845)
(2,0.274084)
(3,1.0309)
(4,0.448005)
(5,0.0841958)
(6,0.603486)
(7,0.36143)
(8,0.061096)
(9,2.07548)
(10,0.72271)
(11,0.287204)
(12,1.15618)
(13,0.345866)
(14,0.446209)
(15,0.0740177)
(16,1.2454)
(17,0.737173)
(18,0.309187)
(19,0.250535)
(20,0.628009)
(21,0.228409)
(22,0.0998563)
(23,0.230796)
(24,0.110218)
(25,0.206598)
(26,0.56572)
(27,0.46654)
(28,0.710515)
(29,0.0242644)
(30,0.0133029)
(31,0.667029)
(32,0.646507)
(33,0.159844)
(34,0.931267)
(35,0.357105)
(36,0.149494)
(37,0.318994)
(38,1.08571)
(39,1.0618)
(40,0.41847)
(41,1.15986)
(42,2.2485)
(43,0.748435)
(44,0.562024)
(45,0.647012)
(46,1.69826)
(47,1.60324)
(48,0.088598)
(49,1.40763)
(50,0.902619)
(51,0.627541)
(52,1.13613)
(53,1.37226)
(54,0.198233)
(55,0.576502)
(56,0.419043)
(57,0.456607)
(58,0.404871)
(59,0.214438)
(60,1.08752)
(61,0.233051)
(62,0.290422)
(63,0.0608396)
(64,0.780158)
(65,0.609633)
(66,1.54391)
(67,2.0416)
(68,0.237252)
(69,0.451634)
(70,0.0339293)
(71,0.963036)
	};
	
\addplot[only marks, color=red] coordinates {
(1,0.71308)
(2,0.296769)
(3,0.661394)
(4,0.378918)
(5,0.419627)
(6,0.0390286)
(7,0.610286)
(8,0.040301)
(9,0.169203)
(10,0.2859)
(11,0.171361)
(12,0.284895)
(13,0.327538)
(14,0.479714)
(15,0.23367)
(16,0.118778)
(17,0.0961612)
(18,0.42836)
(19,0.0823042)
(20,0.168146)
(21,0.0970035)
(22,0.119618)
(23,0.0990226)
(24,0.589986)
(25,0.294236)
(26,0.621241)
(27,0.135501)
(28,0.297047)
(29,0.0223253)
(30,0.284962)
(31,0.385886)
(32,0.286344)
(33,0.487942)
(34,0.415388)
(35,0.0154793)
(36,0.416142)
(37,0.259278)
(38,0.303028)
(39,0.19084)
(40,0.301821)
(41,0.261469)
(42,0.222079)
(43,0.259304)
(44,0.313053)
(45,0.136245)
(46,0.352631)
(47,0.049964)
(48,0.202502)
(49,0.372375)
(50,0.36253)
(51,0.0185583)
(52,0.0131806)
(53,0.130041)
(54,0.0612182)
(55,0.127784)
(56,0.0625813)
(57,0.130617)
(58,0.355359)
(59,0.132776)
(60,0.0267766)
(61,0.59084)
(62,0.199695)
(63,0.0445981)
(64,0.347417)
(65,0.149633)
(66,0.0520898)
(67,0.118598)
(68,0.0511061)
(69,0.28578)
(70,0.331421)
(71,0.259036)
(72,0.218619)
	};
	
\legend{Experimental, Simulation}

\end{axis}
\end{tikzpicture}

%% file: tikz/VelocityCompare.tex
\begin{tikzpicture}
\begin{axis} [
width = 0.5\textwidth,
height=4cm,
xlabel={Image},
ylabel={Velocity in \unit{m/s}},
xlabel style={below},
axis x line=left,
axis y line=left,
xmax=239,
y label style={at={(0,0.5)}},
legend style={at={(0.3,0.2)},anchor=west}
]

\addplot[mark=none, color=blue] coordinates {
(0,0)
(1,0.273178)
(2,0.50512)
(3,0.277966)
(4,0.543847)
(5,0.54834)
(6,0.550073)
(7,0.616119)
(8,0.615189)
(9,0.613218)
(10,0.616445)
(11,0.686912)
(12,0.753403)
(13,0.685254)
(14,0.770767)
(15,0.683729)
(16,0.68218)
(17,0.752622)
(18,0.547877)
(19,0.822526)
(20,0.683071)
(21,0.686842)
(22,0.696103)
(23,0.754889)
(24,0.75504)
(25,0.758301)
(26,0.696103)
(27,0.819928)
(28,0.756651)
(29,0.817415)
(30,0.686793)
(31,0.896129)
(32,0.752746)
(33,0.820154)
(34,0.819436)
(35,0.687404)
(36,0.956747)
(37,0.836402)
(38,0.750907)
(39,0.889581)
(40,0.761309)
(41,0.95676)
(42,0.817501)
(43,0.893714)
(44,0.686733)
(45,0.957439)
(46,0.960342)
(47,0.822097)
(48,0.820154)
(49,0.957363)
(50,0.921895)
(51,0.887795)
(52,0.968354)
(53,0.826028)
(54,1.09838)
(55,0.956737)
(56,0.890959)
(57,0.89398)
(58,0.838993)
(59,1.03485)
(60,1.02925)
(61,0.958174)
(62,0.822555)
(63,1.02236)
(64,1.03167)
(65,1.1006)
(66,0.826039)
(67,1.10139)
(68,0.822097)
(69,1.0286)
(70,1.02644)
(71,0.958863)
(72,0.961931)
(73,0.959542)
(74,0.823558)
(75,1.03253)
(76,1.09737)
(77,1.03452)
(78,0.958937)
(79,0.830063)
(80,1.09822)
(81,0.889879)
(82,1.03204)
(83,1.18035)
(84,0.960788)
(85,1.02994)
(86,0.95955)
(87,0.961454)
(88,1.03002)
(89,1.027)
(90,1.10253)
(91,0.894911)
(92,1.1017)
(93,0.958874)
(94,0.900014)
(95,0.963012)
(96,0.96893)
(97,0.890632)
(98,1.03388)
(99,0.824241)
(100,0.960344)
(101,0.84786)
(102,0.822981)
(103,0.890632)
(104,0.959546)
(105,0.979807)
(106,0.756902)
(107,0.889946)
(108,0.893801)
(109,0.829247)
(110,0.956771)
(111,0.97)
(112,0.90084)
(113,0.965635)
(114,0.894276)
(115,0.965365)
(116,0.963803)
(117,0.965365)
(118,0.9616)
(119,0.961583)
(120,0.971743)
(121,1.0307)
(122,1.03134)
(123,1.03038)
(124,1.1026)
(125,1.03276)
(126,0.963781)
(127,1.10078)
(128,0.962452)
(129,1.03244)
(130,1.03529)
(131,0.893694)
(132,1.03276)
(133,0.965194)
(134,0.892281)
(135,1.03282)
(136,0.966719)
(137,1.02955)
(138,0.962431)
(139,1.03282)
(140,1.10078)
(141,0.966173)
(142,1.1004)
(143,1.03985)
(144,1.03733)
(145,1.10497)
(146,1.03418)
(147,1.0335)
(148,1.10113)
(149,1.1004)
(150,1.04062)
(151,1.03423)
(152,1.1004)
(153,1.03985)
(154,1.03491)
(155,1.1004)
(156,1.04126)
(157,1.03423)
(158,1.10108)
(159,0.96388)
(160,1.10605)
(161,1.03036)
(162,1.03546)
(163,1.03478)
(164,1.03101)
(165,1.03418)
(166,1.03565)
(167,1.04062)
(168,1.03565)
(169,1.1022)
(170,0.96831)
(171,1.10509)
(172,0.969462)
(173,1.03478)
(174,1.03172)
(175,1.03559)
(176,1.03559)
(177,1.03632)
(178,1.03559)
(179,1.03567)
(180,1.03565)
(181,0.967228)
(182,1.04203)
(183,1.03565)
(184,1.10429)
(185,1.03636)
(186,1.03527)
(187,1.04266)
(188,1.03706)
(189,1.04266)
(190,1.03632)
(191,1.10322)
(192,1.10429)
(193,0.971698)
(194,0.969376)
(195,1.10897)
(196,1.03636)
(197,1.03773)
(198,0.975964)
(199,0.968015)
(200,0.968649)
(201,1.1122)
(202,0.96838)
(203,1.0364)
(204,1.04249)
(205,0.971956)
(206,0.96838)
(207,0.975964)
(208,0.8956)
(209,1.03709)
(210,0.897011)
(211,1.03709)
(212,0.829165)
(213,0.974409)
(214,0.910201)
(215,0.900509)
(216,0.904841)
(217,0.90364)
(218,0.828633)
(219,0.835749)
(220,0.973167)
(221,0.900933)
(222,0.899458)
(223,0.927397)
(224,0.901893)
(225,0.829543)
(226,0.834267)
(227,0.831969)
(228,0.704483)
(229,0.694046)
(230,0.635492)
(231,0.628618)
(232,0.556476)
(233,0.491294)
(234,0.486237)
(235,0.437276)
(236,0.276682)
(237,0.348632)
	};
	
\addplot[mark=none, thick, color=red] coordinates {
(0,0)
(1,0.00707107)
(2,0.0339411)
(3,0.175363)
(4,0.304056)
(5,0.369112)
(6,0.468105)
(7,0.538109)
(8,0.580535)
(9,0.611648)
(10,0.636396)
(11,0.648417)
(12,0.671045)
(13,0.695086)
(14,0.7064)
(15,0.721956)
(16,0.727613)
(17,0.721956)
(18,0.712057)
(19,0.711349)
(20,0.712765)
(21,0.713471)
(22,0.724785)
(23,0.734684)
(24,0.740341)
(25,0.747412)
(26,0.758018)
(27,0.778525)
(28,0.813174)
(29,0.837922)
(30,0.839336)
(31,0.825902)
(32,0.814587)
(33,0.816002)
(34,0.818123)
(35,0.819537)
(36,0.827316)
(37,0.837214)
(38,0.842873)
(39,0.846407)
(40,0.852064)
(41,0.863378)
(42,0.873277)
(43,0.873277)
(44,0.871863)
(45,0.872571)
(46,0.871156)
(47,0.868328)
(48,0.868328)
(49,0.876106)
(50,0.887419)
(51,0.897319)
(52,0.905098)
(53,0.917118)
(54,0.93126)
(55,0.941159)
(56,0.942574)
(57,0.936917)
(58,0.937625)
(59,0.94823)
(60,0.956716)
(61,0.955302)
(62,0.953181)
(63,0.960252)
(64,0.975101)
(65,0.987828)
(66,0.992779)
(67,0.990657)
(68,0.989243)
(69,0.987828)
(70,0.982879)
(71,0.972272)
(72,0.967323)
(73,0.965201)
(74,0.965909)
(75,0.968029)
(76,0.968029)
(77,0.975808)
(78,0.988535)
(79,0.995607)
(80,0.992071)
(81,0.98995)
(82,0.997728)
(83,1.01116)
(84,1.00904)
(85,1.00126)
(86,1.00409)
(87,1.01823)
(88,1.03238)
(89,1.03167)
(90,1.0246)
(91,1.01823)
(92,1.0147)
(93,1.00834)
(94,0.9949)
(95,0.976515)
(96,0.96308)
(97,0.95813)
(98,0.951767)
(99,0.94823)
(100,0.949645)
(101,0.951059)
(102,0.950353)
(103,0.942574)
(104,0.928431)
(105,0.905804)
(106,0.891662)
(107,0.888127)
(108,0.881762)
(109,0.87752)
(110,0.881762)
(111,0.893784)
(112,0.912875)
(113,0.929139)
(114,0.938331)
(115,0.94823)
(116,0.955302)
(117,0.96308)
(118,0.97015)
(119,0.980759)
(120,0.989949)
(121,0.995607)
(122,1.0048)
(123,1.01258)
(124,1.01894)
(125,1.02743)
(126,1.03521)
(127,1.03167)
(128,1.01823)
(129,1.00338)
(130,0.995607)
(131,0.987121)
(132,0.97298)
(133,0.963787)
(134,0.968737)
(135,0.982172)
(136,0.994193)
(137,1.00197)
(138,1.00975)
(139,1.02036)
(140,1.03167)
(141,1.0345)
(142,1.03803)
(143,1.04086)
(144,1.04723)
(145,1.05571)
(146,1.06349)
(147,1.06702)
(148,1.07763)
(149,1.08329)
(150,1.08187)
(151,1.07551)
(152,1.0741)
(153,1.07622)
(154,1.06915)
(155,1.05642)
(156,1.0543)
(157,1.06066)
(158,1.06632)
(159,1.06278)
(160,1.05712)
(161,1.05783)
(162,1.05712)
(163,1.05642)
(164,1.04793)
(165,1.03591)
(166,1.02884)
(167,1.03238)
(168,1.03167)
(169,1.03025)
(170,1.04157)
(171,1.05076)
(172,1.0451)
(173,1.0345)
(174,1.03308)
(175,1.03945)
(176,1.03308)
(177,1.01682)
(178,1.0147)
(179,1.02601)
(180,1.04086)
(181,1.0444)
(182,1.04086)
(183,1.04086)
(184,1.04652)
(185,1.05076)
(186,1.05005)
(187,1.0451)
(188,1.03733)
(189,1.04298)
(190,1.04793)
(191,1.05147)
(192,1.05854)
(193,1.0649)
(194,1.06208)
(195,1.05147)
(196,1.04369)
(197,1.04015)
(198,1.03025)
(199,1.00975)
(200,0.9949)
(201,0.991365)
(202,0.998436)
(203,1.00268)
(204,0.99985)
(205,0.997022)
(206,0.992071)
(207,0.987828)
(208,0.977223)
(209,0.965201)
(210,0.951767)
(211,0.946816)
(212,0.942574)
(213,0.932674)
(214,0.924897)
(215,0.925603)
(216,0.93126)
(217,0.924189)
(218,0.905804)
(219,0.889541)
(220,0.883177)
(221,0.885299)
(222,0.88954)
(223,0.886713)
(224,0.881762)
(225,0.876812)
(226,0.840044)
(227,0.764383)
(228,0.714179)
(229,0.711349)
(230,0.707814)
(231,0.673873)
(232,0.624376)
(233,0.585484)
(234,0.546594)
(235,0.506996)
(236,0.463156)
(237,0.422143)
	};
	
\legend{Vehicle Data, IPS Data}

\end{axis}
\end{tikzpicture}

%% file: tables/AccuracyReality.tex
\begin{tabular}{ l c c c}
	& Static & Lines & Circles\\
	Position [cm]  & & & \\
 	Mean  & $1.119578024$ & $1.41932$ & $5.77446$\\
 	Max  & $3.250616575$ & $4.64408$ & $19.4436$\\
 	Std dev & $0.663445809$& $0.804597$ & $3.62245$\\
 	 & & &  \\
 	\hline
 	 & & & \\
 	Orientation [$\degree$] & & & \\
 	Mean & $0.629930184$ & $0.621901$ & $2.10362$\\
 	Max  & $2.248496847$ & $10.4818$ & $12.7505$\\
 	Std dev & $0.516798292$ & $0.438002$ & $1.73485$\\
\end{tabular}

%% file: tikz/EfficiencyPerStep.tex
\begin{tikzpicture} 

\pgfplotsset{ybar stacked, ymin=0, ymax=20, symbolic x coords={3,16, 20}, xtick = {3,16,20}, width= 0.45\textwidth}
  \begin{axis}[bar shift=-8pt,
  legend style={at={(0.5,-0.1)},anchor=north,name = serieA},
  	 y label style={at={(0.08,0.5)}},
     xlabel=Number Vehicles, ylabel= time in \unit{ms}]
    \addplot plot coordinates {(3,4.8)(16,4.4)(20,4.3)};
	\addplot plot coordinates {(3,1.7)(16,2.9)(20,3.4)};
	\addplot plot coordinates {(3,0.1)(16,0.5)(20,0.7) };
	\addplot plot coordinates {(3,0.01)(16,0.02)(20,0.02)};
    \legend{Find Points (Average), Find Vehicles(Average), Match Vehicles (Average), Compute ID and Pose (Average)}
  \end{axis}
  \begin{axis}[bar shift = 8pt,
    legend style = {at = {([yshift = -3mm, xshift=-6mm]serieA.south east)}}, hide axis,]
    \addplot[fill=blue!60] plot coordinates {(3,5.1) (16,5.0)(20,5.6)}; 
	\addplot[fill=red!60] plot coordinates {(3, 1.8) (16, 11.5)(20,12.5)};
	\addplot[fill=yellow!60] plot coordinates { (3,0.1) (16,0.6) (20,0.7)};
	\addplot[fill=black!60] plot coordinates {(3,0.01) (16,0.02) (20,0.02)};
    \legend{Find Points (Worst), Find Vehicles (Worst), Match Vehicles (Worst), Compute ID and Pose (Worst)\textcolor{white}{ni}}
  \end{axis}
  
\end{tikzpicture}

%% file: tikz/EfficiencyAverageCase.tex
\begin{tikzpicture}
\path[use as bounding box] (-0.05,6) rectangle (5, -1);
\begin{axis} [
xbar,
width=0.45\textwidth,
height=7.5cm,
bar width =0.2cm,
axis lines*=left,
symbolic y coords={0,$3-1$,$4-2$,$6-3$,$9-5$,$11-7$,$16-9$,$20-11$, last},
ytick={$3-1$,$4-2$,$6-3$,$9-5$,$11-7$,$16-9$,$20-11$},
ymin={0},
xmin={0},
xmax={32},
xtick distance=5,
xlabel={Time in ms},
y label style={at={(-0.05,0.5)}},
ylabel={\#Vehicles \textminus\textcolor{white}{.} \#Clusters},
legend style={at={(0.8,0.2)},anchor=west}]
\addplot coordinates
{(6.8,$3-1$) (7.3,$4-2$) (7.8,$6-3$) (8.1,$9-5$) (8.3,$11-7$) (8.6,$16-9$) (8.9,$20-11$)};
\addplot coordinates
{(15.9,$3-1$) (15.9,$4-2$) (14.1,$6-3$) (15.7,$9-5$) (17.0,$11-7$) (15.9,$16-9$) (18.3,$20-11$)};

\legend{Mean, Max}

  \draw [dashed,help lines, color=red] (axis cs:20,{last}) -| (axis cs:20,0);
\end{axis}
\end{tikzpicture}

%% file: tikz/EfficiencyWorstCase.tex
\begin{tikzpicture}
\path[use as bounding box] (0,6) rectangle (7, -1);
\begin{axis} [
xbar,
bar width =0.2cm,
width=0.45\textwidth,
height=7.5cm,
axis lines*=left,
symbolic y coords={0,$1$,$6$,$12$,$14$,$16$,$18$,$20$, last},
ytick={$1$,$6$,$12$,$14$,$16$,$18$,$20$},
ymin={0},
xmin={0},
xmax={32},
xtick distance=5,
xlabel={Time in ms},
ylabel={\#Vehicles},
y label style={at={(0,0.5)}},
legend style={at={(0.8,0.2)},anchor=west}]
\addplot coordinates
{(6.5,$1$) (8.7,$6$) (11.5,$12$) (10.6,$14$) (12.7,$16$) (13.0,$18$) (14.8,$20$)};
\addplot coordinates
{(14.1,$1$) (14.4,$6$) (19.0,$12$) (22.0,$14$) (31.5,$16$) (26.2,$18$) (28.6,$20$)};

\legend{Mean, Max}

  \draw [dashed,help lines, color=red] (axis cs:20,{last}) -| (axis cs:20,0);
\end{axis}
   \node[] at (7,3.1) {$98.77\%$};
   \node[] at (7,3.9) {$94.33\%$};
   \node[] at (7,4.7) {$94.38\%$};
   \node[] at (7,5.5) {$87.57\%$};
\end{tikzpicture}

%% file: root.bbl
\begin{thebibliography}{18}
\providecommand{\natexlab}[1]{#1}
\providecommand{\url}[1]{\texttt{#1}}
\providecommand{\urlprefix}{URL }
\expandafter\ifx\csname urlstyle\endcsname\relax
  \providecommand{\doi}[1]{doi:\discretionary{}{}{}#1}\else
  \providecommand{\doi}{doi:\discretionary{}{}{}\begingroup
  \urlstyle{rm}\Url}\fi

\bibitem[{Bradski(2000)}]{opencv_library}
Bradski, G. (2000).
\newblock {The OpenCV Library}.
\newblock \emph{Dr. Dobb's Journal of Software Tools}.

\bibitem[{Chawla et~al.(2010)Chawla, Robins, and Zhang}]{RFID2}
Chawla, K., Robins, G., and Zhang, L. (2010).
\newblock {Object localization using RFID}.
\newblock In \emph{Wireless Pervasive Computing (ISWPC), 2010 5th IEEE
  International Symposium on}, 301--306. IEEE.

\bibitem[{Diab et~al.(2015)Diab, Abel, and Kowalewski}]{DissDiab}
Diab, H., Abel, D., and Kowalewski, S. (2015).
\newblock \emph{{Experimental validation and mathematical analysis of
  cooperative vehicles in a platoon}}.
\newblock Ph.D. thesis, Fachgruppe Informatik, RWTH Aachen University.

\bibitem[{Faessler et~al.(2014)Faessler, Mueggler, Schwabe, and
  Scaramuzza}]{InfraredLeds}
Faessler, M., Mueggler, E., Schwabe, K., and Scaramuzza, D. (2014).
\newblock {A monocular pose estimation system based on infrared LEDs}.
\newblock In \emph{Robotics and Automation (ICRA), 2014 IEEE International
  Conference on}, 907--913.

\bibitem[{Fukuju et~al.(2003)Fukuju, Minami, Morikawa, and Aoyama}]{wireless}
Fukuju, Y., Minami, M., Morikawa, H., and Aoyama, T. (2003).
\newblock {DOLPHIN: An autonomous indoor positioning system in ubiquitous
  computing environment}.
\newblock In \emph{Proceedings IEEE Workshop on Software Technologies for
  Future Embedded Systems. WSTFES 2003}, 53--56. IEEE.

\bibitem[{Ghimire et~al.(2018)Ghimire, Seitz, and Mutschler}]{VLC1}
Ghimire, B., Seitz, J., and Mutschler, C. (2018).
\newblock {Indoor Positioning Using OFDM-Based Visible Light Communication
  System}.
\newblock In \emph{2018 International Conference on Indoor Positioning and
  Indoor Navigation (IPIN)}, 1--8. IEEE.

\bibitem[{Hile and Borriello(2008)}]{Feature1}
Hile, H. and Borriello, G. (2008).
\newblock {Positioning and orientation in indoor environments using camera
  phones}.
\newblock \emph{IEEE Computer Graphics and Applications}, 28(4).

\bibitem[{Ido et~al.(2009)Ido, Shimizu, Matsumoto, and Ogasawara}]{SLAM2}
Ido, J., Shimizu, Y., Matsumoto, Y., and Ogasawara, T. (2009).
\newblock {Indoor navigation for a humanoid robot using a view sequence}.
\newblock \emph{The International Journal of Robotics Research}, 28(2),
  315--325.

\bibitem[{Kloock et~al.(2020)Kloock, Maczijewski, Scheffe, Kampmann,
  Mokhtarian, Kowalewski, and Alrifaee}]{kloock2020cyber}
Kloock, M., Maczijewski, J., Scheffe, P., Kampmann, A., Mokhtarian, A.,
  Kowalewski, S., and Alrifaee, B. (2020).
\newblock {Cyber-Physical Mobility Lab: An Open-Source Platform for Networked
  and Autonomous Vehicles}.
\newblock \emph{arXiv preprint arXiv:2004.10063}.

\bibitem[{Ladd et~al.(2005)Ladd, Bekris, Rudys, Kavraki, and
  Wallach}]{wireless3}
Ladd, A.M., Bekris, K.E., Rudys, A., Kavraki, L.E., and Wallach, D.S. (2005).
\newblock {Robotics-based location sensing using wireless ethernet}.
\newblock \emph{Wireless Networks}, 11(1-2), 189--204.

\bibitem[{Lee and Song(2007)}]{InfraredAreas}
Lee, S. and Song, J.B. (2007).
\newblock {Mobile robot localization using infrared light reflecting
  landmarks}.
\newblock In \emph{2007 International Conference on Control, Automation and
  Systems}, 674--677. IEEE.

\bibitem[{Mautz and Tilch(2011)}]{optical}
Mautz, R. and Tilch, S. (2011).
\newblock {Optical indoor positioning systems}.
\newblock In \emph{Proceedings of the 2011 International Conference on Indoor
  Positioning and Indoor Navigation (IPIN)}.

\bibitem[{Neunert et~al.(2016)Neunert, Bloesch, and Buchli}]{tags}
Neunert, M., Bloesch, M., and Buchli, J. (2016).
\newblock {An open source, fiducial based, visual-inertial motion capture
  system}.
\newblock In \emph{Information Fusion (FUSION), 2016 19th International
  Conference on}, 1523--1530. IEEE.

\bibitem[{Noonan et~al.(2018)Noonan, Rotstein, Geva, and Rivlin}]{SLAM4}
Noonan, J., Rotstein, H., Geva, A., and Rivlin, E. (2018).
\newblock {Vision-Based Indoor Positioning of a Robotic Vehicle with a
  Floorplan}.
\newblock In \emph{2018 International Conference on Indoor Positioning and
  Indoor Navigation (IPIN)}, 1--8. IEEE.

\bibitem[{Olson(2011)}]{tags3}
Olson, E. (2011).
\newblock {AprilTag: A robust and flexible visual fiducial system}.
\newblock In \emph{Robotics and Automation (ICRA), 2011 IEEE International
  Conference on}, 3400--3407. IEEE.

\bibitem[{R{\'a}tosi and Simon(2018)}]{VLC2}
R{\'a}tosi, M. and Simon, G. (2018).
\newblock {Real-Time Localization and Tracking Using Visible Light
  Communication}.
\newblock In \emph{2018 International Conference on Indoor Positioning and
  Indoor Navigation (IPIN)}, 1--8. IEEE.

\bibitem[{Scheffe et~al.(2020)Scheffe, Maczijewski, Kloock, Derks, Kowalewski,
  and Alrifaee}]{scheffe2020}
Scheffe, P., Maczijewski, J., Kloock, M., Derks, A., Kowalewski, S., and
  Alrifaee, B. (2020).
\newblock {Networked and Autonomous Model-scale Vehicles for Experiments in
  Research and Education}.
\newblock 21st IFAC World Congress.

\bibitem[{Yoshino et~al.(2008)Yoshino, Haruyama, and Nakagawa}]{VLC3}
Yoshino, M., Haruyama, S., and Nakagawa, M. (2008).
\newblock {High-accuracy positioning system using visible LED lights and image
  sensor}.
\newblock In \emph{2008 IEEE Radio and Wireless Symposium}, 439--442. IEEE.

\end{thebibliography}
